\documentclass[twocolumn,showpacs,prd]{revtex4}
\usepackage{graphicx}
\usepackage{color}
\usepackage{amssymb}
\newcommand{\ds}{D_s}
\newcommand{\das}{D^\ast}

\newcommand{\dsas}{D_s^\ast}

\newcommand{\kas}{K^\ast}

\newcommand{\dsc}{$D_{s1}(2536)$}
\newcommand{\dsd}{$D_{s1}(2460)$}
\newcommand{\dc}{$D_{1}(2420)$}
\newcommand{\dd}{$D_{1}(2430)$}
\newcommand{\tpz}{${}^{3\!}P_0$}
\newcommand{\tpo}{${}^{3\!}P_1$}

\newcommand{\spo}{${}^{1\!}P_1$}

\newcommand{\ttpt}{$2\,{}^{3\!}P_2$}
\newcommand{\ttpz}{$2\,{}^{3\!}P_0$}

\newcommand{\One}{1\!\!1}
\newcommand{\rse}{\mathcal{R}}
\newcommand{\bes}[1]{j^{#1}_{L_{#1}}}
\newcommand{\han}[1]{h^{(1)#1}_{L_{#1}}}

\newcommand{\tmat}[2]{T_{#1#2}^{(L_{#1},L_{#2})}}
\begin{document}
\title{Quasi-bound states in the continuum: a dynamical coupled-channel
calculation of axial-vector charmed mesons}
\author{Susana Coito}
\email{susana.coito@ist.utl.pt}
\affiliation{Centro de F\'{\i}sica das Interac\c{c}\~{o}es Fundamentais,
Instituto Superior T\'{e}cnico, Technical University of Lisbon,
P-1049-001 Lisboa, Portugal}
\author{George Rupp}
\email{george@ist.utl.pt}
\affiliation{Centro de F\'{\i}sica das Interac\c{c}\~{o}es Fundamentais,
Instituto Superior T\'{e}cnico, Technical University of Lisbon,
P-1049-001 Lisboa, Portugal}
\author{Eef van Beveren}
\email{eef@teor.fis.uc.pt}
\affiliation{Centro de F\'{\i}sica Computacional,
Departamento de F\'{\i}sica, Universidade de Coimbra,
P-3004-516 Coimbra, Portugal}
\date{\today}

\begin{abstract}
Masses and widths of the axial-vector charmed mesons \dc, \dd, \dsc, and \dsd\
are calculated nonperturbatively in the Resonance-Spectrum-Expansion model, by
coupling various open and closed meson-meson channels to the bare $J^P=1^+$
$c\bar{q}$ ($q=u,d$) and $c\bar{s}$
states. The coupling to two-meson channels dynamically mixes and lifts the
mass degeneracy of the spectroscopic \tpo\ and
\spo\ states, as an alternative to the usual spin-orbit splitting. Of
the two resulting $S$-matrix poles in either case, one stays very close to the
energy of the bare state, as a quasi-bound state in the continuum, whereas the
other shifts considerably. This is in agreement with the experimental
observation that the \dc\ and \dsc\ have much smaller widths than one would
naively expect. The whole pattern of masses and widths of the axial-vector
charmed mesons can thus be quite well reproduced with only two free parameters,
one of which being already strongly constrained by previous model calculations.
Finally, predictions for pole positions of radially excited axial-vector
charmed mesons are presented.
\end{abstract}

\pacs{
14.40.Lb, 
13.25.Ft, 
11.80.Gw, 
11.55.Ds, 
12.40.Yx, 
}

\maketitle

\section{Introduction}
\label{intro}
The axial-vector (AV) charmed mesons \dc\ and \dsc\ \cite{PDG2010} have the
puzzling feature that their decay widths are much smaller than one would
expect on the basis of their principal $S$-wave decay modes. Namely, the
\dc\ decays to $D^\ast\pi$ (possibly also in a $D$ wave), with a
phase space of more than 270 MeV, but has a total width of only 20--25
MeV \cite{PDG2010}. On the other hand, the \dsc\ decays to $D^\ast K$
in $S$ and $D$ wave with a phase space of about 30 MeV, resulting in
an unknown tiny width $<\!\!2.3$ MeV, limited by the experimental 
resolution \cite{PDG2010}. The discovery of the missing two AV charmed
mesons, namely the very narrow \dsd\ and the very broad
\dd, first observed by CLEO \cite{PRD68p032002} and Belle
\cite{PRD69p112002}, respectively, completed an even more confusing
picture. While the tiny width of the \dsd\ can be easily understood, since
this meson lies underneath its lowest Okubo-Zweig-Iizuka--allowed (OZIA) and
isospin-conserving decay threshold, the huge \dd\ width, in $D^\ast\pi$, is
in sharp contrast with that of the \dc. Moreover, the \dsd\ lies 76 MeV below
the \dsc, whereas the \dc\ and \dd\ are almost degenerate in mass, if one
takes the central value of the latter resonance.

Quark potential models, with standard spin-orbit splittings, fail dramatically
in reproducing this pattern of masses. For instance, in the relativized quark
model \cite{PRD32p189} the $c\bar{s}$ state that is mainly \tpo\ comes 
out at 2.57 GeV, assuming the already then well-established \dsc\ to be
mostly \spo, though with a very large mixing between \tpo\ and \spo. Reference
\cite{PRD32p189} similarly predicted a too high mass for the dominantly \tpo\
state in the $c\bar{q}$ ($q=u,d$)
sector, viz. 2.49~GeV. In the chiral quark model for heavy-light systems of
Ref.~\cite{PRD64p114004}, the result for the mainly \tpo\
$c\bar{q}$ state is also 2.49~GeV, while the discrepancy
is even worse in the $c\bar{s}$ sector, with a prediction of 2.605~GeV for the
mostly \tpo\ state, now with a small mixing in both sectors.

More recently and after the discovery of the \dsd\ (and \dd), chiral Lagrangians
for heavy-light systems (see e.g.\
Refs.~\cite{PRD68p054024,PLB599p55,MPLA19p2083,PRD72p034006}) have
been employed in order to understand the masses of the AV charmed mesons, in
particular the mass splittings with respect to the vector (V) mesons with charm
$D_s^\ast$ and $D^\ast$, respectively. Reference~\cite{PLB599p55} analyzed in
detail the curious experimental \cite{PDG2010} observation that the AV-V mass
difference is considerably larger in the charm-nonstrange sector than in the
charm-strange one, which is not predicted by typical quark potential models
\cite{PRD32p189,PRD64p114004}. The same discrepancy applies to the
scalar-pseudoscalar mass difference in either sector \cite{PDG2010,PLB599p55}.
In Ref.~\cite{PLB599p55}, the problem was tackled by calculating chiral loop
corrections, but the result turned out to be exactly the opposite of what is
needed to remove or alleviate the discrepancy.

An alternative approach to the AV charmed mesons is by trying to generate them
as dynamical resonances in chiral unitary theory \cite{EPJA33p119}. Indeed, in
the latter paper, describing AV mesons in other flavor sectors as well,
several charmed resonances were predicted, including the \dc, \dd, \dsc, and
\dsd, with reasonable results, though the $c\bar{q}$ states
came out about 100 MeV off. However, dynamical generation of mesonic resonances,
including the ones that are commonly thought to be of a normal 
quark-antiquark
type, may give
rise to interpretational difficulties, besides predicting several genuinely
exotic and so far unobserved states \cite{EPJA33p119}.
Dynamically generated AV charmed as well as bottom mesons can be found in
Ref.~\cite{PLB647p133},
too.

Finally, in
Ref.~\cite{PRD77p074017} a coupled-channel calculation of positive-parity
$c\bar{s}$ and $b\bar{s}$ was carried out in a chiral quark model, similar
to our approach in its philosophy, and with results for the \dsc\ and \dsd\
close to the present ones (also see below).

\section{Resonance-Spectrum Expansion}
In the present
paper,
we employ the Resonance-Spectrum Expansion (RSE)
to describe the AV charmed mesons. The RSE model has been developed for
meson-meson (MM) scattering in non-exotic channels, whereby the intermediate
state is described via an infinite tower of $s$-channel $q\bar{q}$ states
\cite{AOP324p1620}. For the spectrum of the latter, in principle
any confinement potential can be employed, but in practical applications, a
harmonic oscillator (HO) with constant frequency has been used, with excellent
results \cite{AOP324p1620}. Recent applications of the RSE concern the
$\phi(2170)$ \cite{PRD80p094011} and $X(3872)$ \cite{1008.5100} resonances.

In order to account for the two possible spectroscopic channels \tpo\ and \spo\
contributing to a $J^P=1^+$ state with undefined $C$-parity, we couple both
$q\bar{q}$ channels to the most important meson-meson channels.
The resulting fully off-energy-shell RSE $T$ matrix reads
\cite{AOP324p1620,PRD80p094011,1008.5100}
\begin{eqnarray}
\lefteqn{\tmat{i}{j}(p_i,p'_j;E)=-2\lambda^2\sqrt{\mu_ip_ir_0}\,\bes{i}(p_ir_0)
\times} \nonumber \\
&&\hspace*{-10pt}\sum_{m=1}^{N}\rse_{im}\left\{[\One-\Omega\,\mathcal{R}]^{-1}
\right\}_{\!mj}\bes{j}(p'_jr_0)\,\sqrt{\mu_jp'_jr_0} \; ,
\label{tmat}
\end{eqnarray}
with the diagonal loop function
\begin{equation}
\Omega_{ij}(k_j)=-2i\lambda^2\mu_jk_jr_0\,\bes{j}(k_jr_0)\,\han{j}(k_jr_0)\,
\delta_{ij}\;, 
\label{omega}
\end{equation}
and the RSE propagator
\begin{equation}
\mathcal{R}_{ij}(E)=\sum_{S=0,1}\sum_{n=0}^{\infty}
\frac{g^i_{(S,n)}g^j_{(S,n)}}{E-E_n^{(S)}}\;.
\label{rse}
\end{equation}
Here, $\lambda$ is an overall coupling,
$r_0$ is the average distance for decay via \tpz\ quark-pair creation,
$E_n^{(S)}$ is the discrete energy of the $n$-th recurrence in the $q\bar{q}$
channel with spin $S$, $g^i_{(S,n)}$ is the corresponding coupling to the
$i$-th MM channel, $\mu_i$ the reduced mass for this channel, $p_i$ the
off-shell relative momentum, $L_i$ the orbital angular momentum, and
$\bes{i}$ and $\han{j}(k_jr_0)$ the spherical Bessel and Hankel
functions of the first kind, respectively. Note that $\mu_i$, $p_i$, and the
on-energy-shell relative momentum $k_i$ are defined relativistically.
Also notice that the infinite sum over the higher recurrences converges very
fast, so that it can be truncated after 20 terms in practical calculations.
The $S$ matrix is finally given by
$S^{(L_i,L_j)}_{ij}(E)=1+2i\tmat{i}{j}(k_i,k_j;E)$.

\section{OZI-allowed channels for AV charmed mesons}
\label{ozi}
Now we describe the physical AV charmed resonances by coupling bare \tpo\ and
\spo\ $c\bar{q}$, 
$c\bar{s}$
channels to all OZI-allowed ground-state pseudoscalar-vector
(PV) and vector-vector (VV) channels. It is true that there are also relevant
\begin{table}[b]
\caption{Included meson-meson channels for \dc\ and \dd, with ground-state
couplings squared
\cite{ZPC21p291}, 
orbital angular momenta, and thresholds in MeV. For $\eta$ and $\eta'$, a
pseudoscalar mixing angle of $37.3^\circ$ \cite{PRD80p094011} is used.}
\centering
\begin{tabular}{ccccc}
\hline \hline &&& \\[-11pt]
Channel & $\left(\tilde{g}^i_{(S=1,n=0)}\right)^2$ &
$\left(\tilde{g}^i_{(S=0,n=0)}\right)^2$ & $L$ & Threshold \\ [0.5ex]
\hline
$\das\pi$    & 0.02778  &   0.01389  &    0 & 2146\\
$\das\pi$    & 0.03472  &   0.06944  &    2 & 2146\\
$\das\eta$   & 0.00524  &   0.00262  &    0 & 2556\\
$\das\eta$   & 0.00655  &   0.01310  &    2 & 2556 \\
$\dsas K$    & 0.01852  &   0.00926  &    0 & 2608\\
$\dsas K$    & 0.02315  &   0.04630  &    2 & 2608\\
$D\rho$      & 0.02778  &   0.01389  &    0 & 2643\\
$D\rho$	     & 0.03472  &   0.06944  &    2 & 2643\\
$D\omega$    & 0.00926  &   0.00463  &    0 & 2650\\
$D\omega$    & 0.01157  &   0.02315  &    2 & 2650\\
$\das\rho$   & 0        &   0.01389  &    0 & 2784\\
$\das\rho$   & 0.01042  &   0.06944  &    2 & 2784\\
$\das\omega$ & 0        &   0.00463  &    0 & 2791\\
$\das\omega$ & 0.03472  &   0.02315  &    2 & 2791\\
$\ds\kas$    & 0.01852  &   0.00926  &    0 & 2862\\
$\ds\kas$    & 0.02315  &   0.04630  &    2 & 2862\\
$\das\eta'$  & 0.00402  &   0.00201  &    0 & 2996\\
$\das\eta'$  & 0.00502  &   0.01004  &    2 & 2996\\
$\dsas\kas$  & 0        &   0.00926  &    0 & 3006\\
$\dsas\kas$  & 0.06944  &   0.04630  &    2 & 3006\\
\hline \hline
\end{tabular}
\label{cn}
\end{table}
\begin{table}[t]
\caption{As Table~\ref{cn}, but now for  \dsc\ and \dsd.}
\centering
\begin{tabular}{ccccc}
\hline \hline &&& \\[-11pt]
Channel & $\left(\tilde{g}^i_{(S=1,n=0)}\right)^2$ &
 $\left(\tilde{g}^i_{(S=0,n=0)}\right)^2$
& L & Threshold \\[0.5ex] 
\hline
$\das K$     & 0.03704 & 0.01852 & 0 & 2504\\
$\das K$     & 0.04630 & 0.09259 & 2 & 2504\\
$\dsas\eta$  & 0.00803 & 0.00402 & 0 & 2660\\
$\dsas\eta$  & 0.01004 & 0.02009 & 2 & 2660\\
$D\kas$	     & 0.03704 & 0.01852 & 0 & 2761\\
$D\kas$	     & 0.04630 & 0.09259 & 2 & 2761\\
$\das\kas$   & 0       & 0.01852 & 0 & 2902\\
$\das\kas$   & 0.01389 & 0.09259 & 2 & 2902\\
$\ds\phi$    & 0.01852 & 0.00926 & 0 & 2988\\
$\ds\phi$    & 0.02315 & 0.04630 & 2 & 2988\\
$\dsas\eta'$ & 0.01048 & 0.00524 & 0 & 3069\\
$\dsas\eta'$ & 0.01310 & 0.02621 & 2 & 3069\\
$\dsas\phi$  & 0       & 0.00926 & 0 & 3132\\
$\dsas\phi$  & 0.06944 & 0.04630 & 2 & 3132\\
\hline \hline
\end{tabular}
\label{cs}
\end{table}
pseudoscalar-scalar (PS) channels (in $P$-wave), most notably $Df_0(600)$
and $D_0^\ast(2400)\pi$ \cite{PDG2010} in the AV $c\bar{q}$
case, and $DK_0^\ast(800)$ for $c\bar{s}$. These will contribute to the
observed \cite{PDG2010} $D\pi\pi$ and $D\pi K$ decay modes, respectively. Now,
we have recently developed \cite{1008.5100} an algebraic procedure to deal with
resonances in asymptotic states whilst preserving unitarity. However, the huge 
widths of the $D_0^\ast(2400)$, $f_0(600)$, and $K_0^\ast(800)$ resonances may
lead to fine sensitivities that will tend to obscure the point we want to make,
apart from the fact that there will also be nonresonant contributions to the
$D\pi\pi$ and $D\pi K$ final states. So we restrict ourselves to the open and
closed PV and VV channels in the present investigation, but we shall further
discuss this issue below. The here included channels for
$c\bar{q}$ and $c\bar{s}$ are given in Tables~\ref{cn} and
\ref{cs}, respectively, together with the corresponding orbital angular momenta,
threshold energies, and ground-state couplings squared
$(\tilde{g}^i_{(S=1(0),n=0})^2$,
where $S=1(0)$ refers to the \tpo\ (\spo) quark-antiquark component.
In Appendix~\ref{couplings}, we show in more detail how the ground-state 
coupling constants in Tables~\ref{cn} and \ref{cs} depend on the isospin
and $J^{PC}$ quantum numbers of the various meson-meson channels.
The latter
squared couplings,
computed in the very general framework of Ref.~\cite{ZPC21p291},
must be multiplied
by $(n+1)/4^n$ for $L=0$ and by $(2n/5+1)/4^n$ for $L=2$, so as to obtain the
couplings for the radial recurrences $n$ in the RSE sum of Eq.~(\ref{rse}).
Note that the scheme of Ref.~\cite{ZPC21p291} employs overlaps of HO
wave functions for the original $q\bar{q}$ pair, the \tpz\ $q\bar{q}$ pair
created out of the vaccuum, and the outgoing mesons. This allows to rigorously
calculate the coupling constants of all excited states as well, in contrast
with approaches using combinations of Clebsch-Gordan coefficients only.
Nevertheless, our ground-state couplings are identical to the usual ones in
practically all situations, including the present one.
Finally,
a subthreshold suppression of closed channels is used just as in
Ref.~\cite{PRD80p094011}.

The energies of the bare AV $c\bar{q}$ and $c\bar{s}$
states we determine, as in previous work (see e.g.\
Refs.~\cite{PRD80p094011,1008.5100}), from an HO spectrum. The corresponding
constant oscillator frequency and the constituent masses of the charmed,
strange, and nonstrange quarks are also kept completely unchanged at the values
$\omega\!=\!190$ MeV, $m_c\!=\!1562$ MeV, $m_s\!=\!508$ MeV, and $m_n\!=\!406$
MeV \cite{PRD80p094011,1008.5100}. This yields masses of $2443$ MeV and $2545$
MeV for the bare AV $c\bar{q}$ and $c\bar{s}$ states,
respectively, which are very close to values found in typical single-channel
quark models \cite{PRD32p189,PRD64p114004}. 

\section{Quasi-bound states in the continuum and other poles}
\label{poles}
\begin{figure}[t]
\mbox{} \\[-30pt]
\hspace*{-30pt}
\resizebox{!}{470pt}{\includegraphics{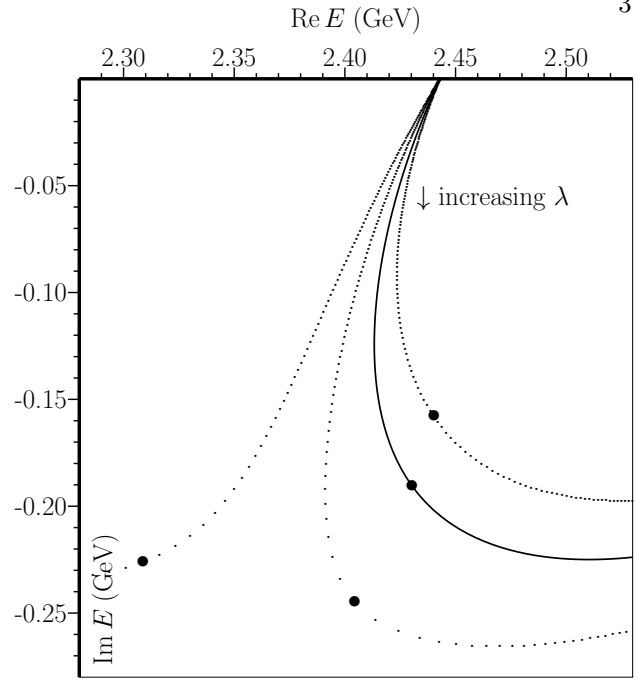}}
\mbox{} \\[-205pt]
\caption{\dd\ pole trajectories as a function of $\lambda$, for
$r_0=$ 3.2--3.5 GeV$^{-1}$ (left to right). Solid curve and bullets correspond
to $r_0=$ 3.40~GeV$^{-1}$ and $\lambda=1.30$, respectively.}
\label{donesoft}
\end{figure}

\begin{figure}[t]
\mbox{} \\[-35pt]
\hspace*{-30pt}
\resizebox{!}{470pt}{\includegraphics{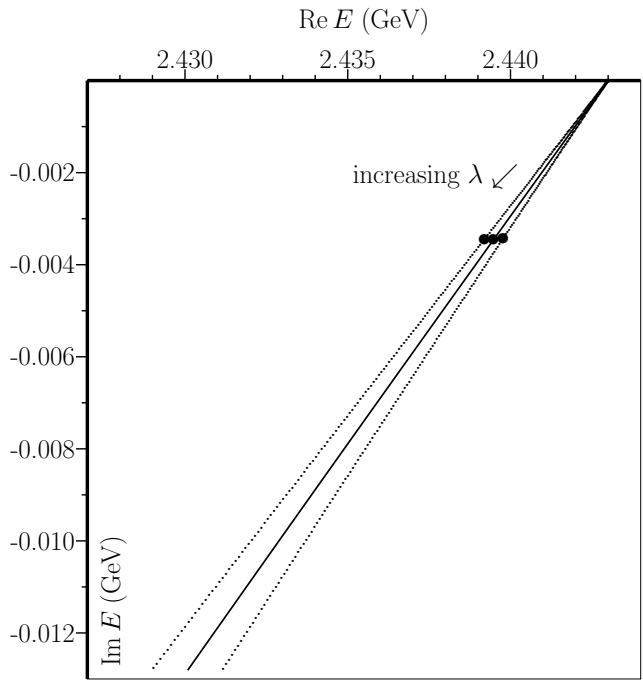}}
\mbox{} \\[-205pt]
\caption{\dc\ pole trajectories as a function of $\lambda$, for
$r_0=$ 3.3--3.5 GeV$^{-1}$ (left to right). Solid curve and bullets correspond
to $r_0=$ 3.40~GeV$^{-1}$ and $\lambda=1.30$, respectively.}
\label{donehard}
\end{figure}

Next we search for poles in the $S$ matrix. Starting with the
$c\bar{q}$ case, we choose $r$ in the range
3.2--3.5~GeV$^{-1}$ (0.64--0.70~fm), which is in
between the values of 2.0~GeV$^{-1}$ \cite{1008.5100} for an AV $c\bar{c}$
system and 4.0~GeV$^{-1}$ \cite{PRD80p094011} for vector $s\bar{s}$ states. In
Fig.~\ref{donesoft}, we plot several pole trajectories in the complex E plane
as a function of the overall coupling $\lambda$.
We see that this pole rapidly acquires a large imaginary part, whereas the
real part changes considerably less, especially in the range 
$r_0=3.3$--$3.5$ GeV$^{-1}$, making it a good candidate for the broad \dd\
resonance. For $\lambda=1.30$ and $r_0=3.40$
GeV$^{-1}$, the pole comes out at
$(2430-i\times191)$~MeV, being thus
fine-tuned to the experimental mass and width \cite{PDG2010}.
However, there should be another pole in the $S$ matrix, since there are 2
quark-antiquark channels and more than 2 MM channels. From the structure of 
the $T$ matrix in Eqs.~(\ref{tmat}--\ref{rse}), one can algebraically show
that the number of poles for each bare state is equal to
min$(N_{q\bar{q}},N_{MM})$, besides possible poles of a purely dynamical
nature. Indeed, another pole originating from the bare $c\bar{q}$ state is
encountered, with its trajectories depicted in Fig.~\ref{donehard}. Quite
remarkably, this pole moves very little, acquiring an imaginary part that
is a factor 55 smaller than in the \dd\ case, for the values
$\lambda=1.30$ and $r_0=3.40$~GeV$^{-1}$
(see solid lines and bullets in both figures). So this resonance, with a pole
position of $(2439-i\times3.5)$ MeV, almost decouples from the
only open OZIA MM channel \cite{EPJC32p493}, viz.\ $D^\ast\pi$, representing a
quasi-bound state in the continuum (QBSC) \cite{PLB647p133}.
Moreover, it is a good candidate for the \dc, though
its width of roughly 7~MeV is somewhat too small and its mass 16 MeV too high.
These minor discrepancies may be due to the neglect of the PS channels, with
broad resonances in the final states, as suggested above.

\begin{figure}[hb]
\mbox{} \\[-15pt]
\hspace*{-35pt}
\resizebox{!}{470pt}{\includegraphics{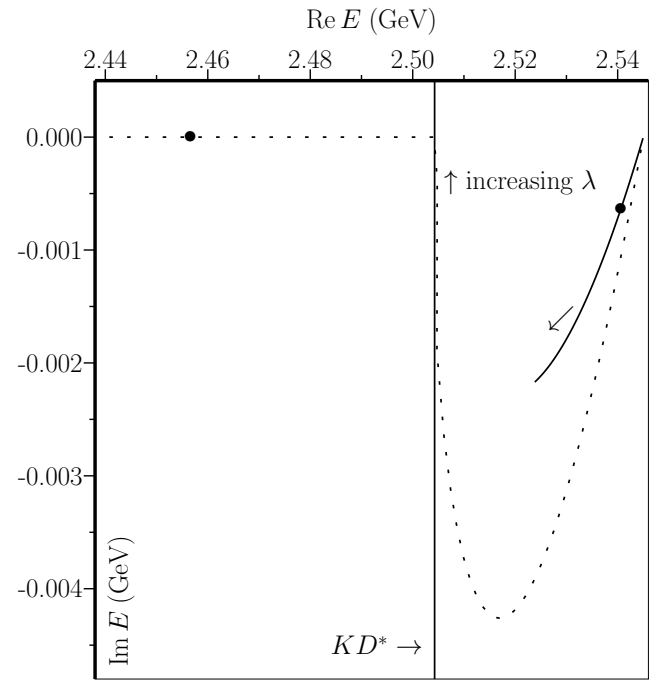}}
\mbox{} \\[-215pt]
\caption{\dsd\ (dashed) and \dsc\ (solid) pole trajectories as a
function of $\lambda$, for $r_0=$ 3.12 GeV$^{-1}$. Bullets correspond to
$\lambda=1.19$; vertical line shows $D^\ast K$ threshold.}
\label{dsonesofthard}
\end{figure}

Nevertheless, these encouraging results might be partly due to a fortuitous
choice of the parameters $\lambda$ and
$r_0$.
Therefore, we now check the
$c\bar{s}$ system, 
thereby scaling $r_0$ and $\lambda$ with the square root of the reduced quark
mass (see Ref.~\cite{EPJC32p493}, Eq.~(13)), so as to
respect flavor independence of our equations,
which yields the $c\bar{s}$ values $r_0=3.12$~GeV$^{-1}$ and $\lambda=1.19$.
The ensuing $c\bar{s}$ pole trajectories are depicted in
Fig.~\ref{dsonesofthard}, but
now for $r_0=3.12$~GeV$^{-1}$ only. Thus, for
$\lambda=1.19$, the strongly
coupling state comes out at 2452 MeV, i.e., only 7.5 MeV below
the \dsd\ mass, with a vanishing width, as the pole ends up below the lowest
OZIA channel. As for the $c\bar{s}$ QBSC, it indeed shifts very little from
the bare state, settling at $(2540 - i\times0.7)$~MeV, i.e., only 5 MeV
above the \dsc\ mass, and having a width fully compatible with experiment
\cite{PDG2010}.

Besides the above ground-state AV charmed mesons, the present model of course
also predicts higher recurrences of these resonances. However, due caution is 
necessary so as to account for the most relevant open and closed decay channels
at the relevant energy scales. Now, the first radially excited HO levels of the
\tpo/\spo\ $c\bar{n}$ and $c\bar{s}$ states lie at 2823~MeV and 2925~MeV,
respectively,
which allows the corresponding resonances to be reasonably described by the
channels included in Tables~\ref{cn}, \ref{cs}. Thus, we find again 4 poles,
tabulated in Table~III, together with those of the ground-state
\begin{table}[ht]
\caption{Poles of ground-state ($n\!=\!0$) and first
radially-excited ($n\!=\!1$) AV charmed mesons. Parameters: $\lambda=1.30$
(1.19) and $r_0=$ 3.40 (3.12) GeV$^{-1}$, for $c\bar{q}$ ($c\bar{s}$) states.}
\centering
\begin{tabular}{ccl}
\hline \hline && \\[-11pt]
Quark Content & Radial Excitation & Pole in MeV \\[0.5ex] 
\hline
$c\bar{q}$  & $0$ & $2439 - i\times3.5$     \\
$c\bar{q}$  & $0$ & $2430 - i\times191$     \\
$c\bar{s}$  & $0$ & $2540 - i\times0.7$     \\
$c\bar{s}$  & $0$ & $2452 - i\times0.0$     \\
$c\bar{q}$  & $1$ & $2814 - i\times7.8$     \\
$c\bar{q}$  & $1$ & $2754 - i\times47.2$    \\
$c\bar{s}$  & $1$ & $2915 - i\times6.7$     \\
$c\bar{s}$  & $1$ & $2862 - i\times25.7$    \\
\hline \hline
\end{tabular}
\label{recurrences}
\end{table}
AV charmed mesons. For the radially excited states, we observe a similar
pattern as for the ground states, namely two poles that remain close
to the bare HO levels, whereas two other poles shift considerably. Note,
however, that the difference is not as dramatic as in the $n=0$ case.
This may be due to the fact that several decay channels are open now. As
for a possible observation of the here predicted $2\,P_1$ states, no
experimental candidates have been reported so far. Namely, in the nearby
$c\bar{q}$ mass region, the two listed \cite{PDG2010} resonances $D(2600)$
and $D(2750)$ \cite{PDG2010} both decay to $D^\ast\pi$ and $D\pi$, which
excludes an AV assignment.

Concerning the $c\bar{s}$ sector, the only listed \cite{PDG2010} state
around 2.8--2.9~GeV is the $D^\ast_{sJ}(2860)$
\cite{PRD80p092003},
with natural parity and so not an
AV, decaying to $D^\ast K$ and $DK$, which makes it a good candidate for the
\ttpt\ state, possibly overlapped by the \ttpz\ \cite{PRD81p118101}. Note that
the lower of our two predicted $2\,P_1$ resonances also practically coincides
with the $D^\ast_{sJ}(2860)$, both in mass and width. This may be a further
indication that the $D^\ast_{sJ}(2860)$ structure corresponds to more than one
resonance only.

To conclude this section, we study --- for the $c\bar{q}$ system --- the
dependence of the lowest-lying poles on the number of included quark-antiquark
and MM channels. In Table~\ref{poletests}, besides the \dc\ and \dd\
\begin{table}[ht]
\caption{Poles of AV $c\bar{q}$ mesons, for different sets of
included channels. Parameters: $\lambda=1.30$, $r_0=3.40$~GeV$^{-1}$.}
\centering
\begin{tabular}{cccc}
\hline \hline &&& \\[-11pt]
$c\bar{q}$ channels & MM channels & Pole 1 (MeV) & Pole 2 (MeV)
\\[0.5ex] 
\hline
\tpo+\spo  & $20$ & $2430 - i\times191$      & $2439 - i\times3$  \\
\tpo+\spo  & $2$  & $2402 - i\times36$\,\,\, & $2441 - i\times1$  \\
\tpo+\spo  & $1$  & $2431 - i\times39$\,\,\, &       -            \\
\tpo       & $20$ & $2409 - i\times65$\,\,\, &       -            \\
\spo       & $20$ & $2425 - i\times96$\,\,\, &       -            \\
\hline \hline
\end{tabular}
\label{poletests}
\end{table}
poles resulting from the full calculation, with the 20 MM channels from
Table~\ref{cn}, we first give the pole positions for the cases that only 2
($D^\ast\pi$, $L=0,2$) or 1 ($D^\ast\pi$, $L=0$) MM channels are included.
The last two poles then correspond to calculations with the full 20 MM
channels but only one quark-antiquark channel, viz.\ \tpo\ or \spo. Notice
that only one pole is found when the number of quark-antiquark or MM channels
is equal to 1. This confirms our above conjecture that the number of poles
for each bare HO level is given by min$(N_{q\bar{q}},N_{MM}$).

\section{Summary and conclusions}
\label{summary}
In the foregoing,
we
have managed
to rather accurately reproduce the masses and widths of the \dc, \dd, \dsc,
and \dsd\ with only 2 free parameters, one of which is already constrained
by previous model calculations, as well as by reasonable estimates for the
size of these mesons.
Crucial is the approximate decoupling from the continuum of one combination
of \tpo\ and \spo\ components, which amounts to a mixing angle close to
$35^\circ$.
Namely, if we express a QBSC as
$|\mbox{QBSC}\rangle=-\sin\theta\,|$\tpo$\rangle+\cos\theta\,|$\spo$\rangle$,
it decouples from the $L\!=\!0$ $D^\ast\pi$ channel (for $c\bar{q}$) or
$D^\ast K$ channel (for $c\bar{s}$), if
$\theta=\arccos\sqrt{2/3}\approx35.26^\circ$ (see Tables~\ref{cn}, \ref{cs}).
Inclusion of the other, practically all closed, channels apparently changes
the picture only slightly in our formalism.
This
result
is in full agreement with
the findings in Ref.~\cite{PRD77p074017}. However, in the present approach this
particular mixing \cite{1105.6025} comes out as a completely dynamical result,
and is not chosen by us beforehand. Moreover, the bare-mass degeneracy of
\tpo\ and \spo\ states is adequately lifted via the decay couplings in
Tables~\ref{cn} and \ref{cs}, dispensing with the usual $\vec{S}\cdot\vec{L}$
splitting. Also note that the occurrence of (approximate) bound states in the
continuum for AV charmed mesons had already been conjectured by two of us
\cite{EPJC32p493}, based on more general arguments.

The puzzling discrepancy between the AV-V mass splittings in the
$c\bar{q}$ and $c\bar{s}$ sectors is resolved in our
calculation by dynamical, nonperturbative coupled-channel effects. A similar
phenomenon we have observed before \cite{PRL91p012003} for the
$D_0^\ast$(2300--2400) \cite{PDG2010} resonance, and may be related to an
effective Adler-type zero \cite{PR397p257} in the $D^\ast\pi$ and $D\pi$
channels in the AV and scalar $c\bar{n}$ cases, respectively, owing to the
small pion mass.

Summarizing, we have reproduced the whole pattern of masses and widths of the
AV charmed mesons dynamically, by coupling the most important open and closed
two-meson channels to bare $c\bar{q}$ and $c\bar{s}$
states containing both \tpo\ and \spo\ components. The dynamics of the
coupled-channel equations straightforwardly leads to one pair of strongly
shifted states and another pair of QBSCs. Ironically, the state that shifts
most in mass, namely the \dsd, ends up as the narrowest resonance. This 
emphasizes the necessity \cite{PTP125p581} to deal with unquenched meson
spectroscopy in a fully nonperturbative framework.

One might argue that these conclusions will depend on the specific model
employed. Admittedly, our numerical results could change somewhat if slightly
different bare masses for the AV charmed mesons were chosen,  non-$S$-wave
decay channels were included as well, or a different scheme was used to
calculate the decreasing couplings of the higher recurrences. Nevertheless,
we are convinced the bulk of our results will not change, most notably
the appearance of QBSCs and the large shifts of their partner states, as the
almost inevitable consequence of exact nonperturbative coupled-channel
dynamics.

\begin{acknowledgments}
This work was supported in part by the {\it Funda\c{c}\~{a}o para a
Ci\^{e}ncia e a Tecnologia} \/of the {\it Minist\'{e}rio da Ci\^{e}ncia,
Tecnologia e Ensino Superior} \/of Portugal, under contract 
CERN/FP/116333/2010 and grant SFA-2-91/CFIF.
\end{acknowledgments}

\appendix
\section{Three-meson couplings}
\label{couplings}
The ground-state couplings in Tables~\ref{cn} and \ref{cs} are obtained
by multiplying the isospin recouplings given in Table~\ref{isospin} with the 
$J^{PC}$ couplings in Table~\ref{jpc}, for an OZIA process $M_A\to M_B+M_C$
based on \tpz\ $q\bar{q}$ creation \cite{ZPC21p291}. For clarity, we 
represent here all couplings by rational numbers. Note that $\eta_n$ and
$\eta_s$ in Table~\ref{isospin} stand for the pseudoscalar $I=0$ states 
$(u\bar{u}+d\bar{d})/\sqrt{2}$ and $s\bar{s}$, respectively. Then, we get
the couplings to the physical $\eta$ and $\eta^\prime$ mesons 
by applying a mixing angle --- in the flavor basis --- of $41.2^\circ$,
as in Ref.~\cite{PRL91p012003}, 2nd paper. For the $\omega$ and $\phi$ we
assume ideal mixing.
\begin{table}[ht]
\caption{Squared isospin recouplings for the 3-meson process
$M_A\to M_B+M_C$, with $M_A=c\bar{s}$ or $c\bar{q}$.}
\centering
\begin{tabular}{cccc}
\hline \hline && \\[-11pt]
$\;\;\;M_A\;\;\;$ & $\;\;\;M_B\;\;\;$ & $\;\;\;M_C\;\;\;$ &
$\;\;\;g_I^2\;\;\;$ \\[0.5ex]
\hline
$D_{s1}$ & $D_s,D_s^\ast$ & $\eta_s,\phi$ & $1/3$ \\
$D_{s1}$ & $D,D^\ast$     & $K,K^\ast$    & $2/3$ \\
\hline
$D_{1}$  & $D_s,D_s^\ast$ & $K,K^\ast$      & $1/3$ \\
$D_{1}$  & $D,D^\ast$     & $\pi,\rho$      & $1/2$ \\
$D_{1}$  & $D,D^\ast$     & $\eta_n,\omega$ & $1/6$ \\
\hline \hline
\end{tabular}
\label{isospin}
\end{table}
\begin{table}[ht]
\caption{Squared ground-state coupling constants for the 3-meson
process $M_A\to M_B+M_C$, with $J^{PC}(M_A)=1^{+\pm}$, and $M_A$, $M_B$
belonging to the lowest pseudoscalar or vector nonet.}
\centering
\begin{tabular}{cccccc}
\hline\hline &&&& \\[-11pt]
$J^{PC}(M_A)$ & $J^{PC}(M_B)$ & $J^{PC}(M_C)$ & $L_{M_B M_C}$ & $S_{M_B M_C}$ &
$g^2_{(n=0)}$
\\[0.5ex]
\hline
$1^{++}$ & $0^{-+}$ & $1^{--}$ & $0$ & $1$ & $1/18$\\
$1^{++}$ & $0^{-+}$ & $1^{--}$ & $2$ & $1$ & $5/72$\\
$1^{++}$ & $1^{--}$ & $1^{--}$ & $0$ & $1$ & $0$\\
$1^{++}$ & $1^{--}$ & $1^{--}$ & $2$ & $2$ & $5/24$\\
$1^{+-}$ & $0^{-+}$ & $1^{--}$ & $0$ & $1$ & $1/36$\\
$1^{+-}$ & $0^{-+}$ & $1^{--}$ & $2$ & $1$ & $5/36$\\
$1^{+-}$ & $1^{--}$ & $1^{--}$ & $0$ & $1$ & $1/36$\\
$1^{+-}$ & $1^{--}$ & $1^{--}$ & $2$ & $1$ & $5/36$\\
\hline \hline
\end{tabular}
\label{jpc}
\end{table}

\newcommand{\pubprt}[4]{#1 {\bf #2}, #3 (#4)}
\newcommand{\ertbid}[4]{[Erratum-ibid.~#1 {\bf #2}, #3 (#4)]}
\def\AP{Ann.\ Phys.}
\def\PAN{Phys.\ Atom.\ Nucl.}
\def\PLB{Phys.\ Lett.\ B}
\def\PRD{Phys.\ Rev.\ D}
\def\PRL{Phys.\ Rev.\ Lett.}

\end{document}